\documentclass{PoS}
\usepackage{bm}
\usepackage{comment}

\title{A lattice formulation of the Atiyah-Patodi-Singer index\footnote{The original title of the talk was ``Atiyah-Patodi-Singer index theorem on a lattice.''}}

\ShortTitle{Atiyah-Patodi-Singer index theorem on a lattice}

\author{
		Hidenori Fukaya$^a$\thanks{E-mail:
hfukaya@het.phys.sci.osaka-u.ac.jp}\  , 
        \speaker{Naoki Kawai}$^a$\thanks{E-mail:
nkawai@het.phys.sci.osaka-u.ac.jp}\  ,
        Yoshiyuki Matsuki$^{a}$\thanks{E-mail: ymatsuki@het.phys.sci.osaka-u.ac.jp}\ \,
        Makito Mori$^{a}$\thanks{E-mail:
m-mori@het.phys.sci.osaka-u.ac.jp}\  ,
		Katsumasa Nakayama$^{{a}{b}}$\thanks{E-mail: katsumasa@het.phys.sci.osaka-u.ac.jp}\  ,
        Tetsuya Onogi$^{a}$\thanks{E-mail: onogi@het.phys.sci.osaka-u.ac.jp}\ \ , and
        Satoshi Yamaguchi$^{a}$\thanks{E-mail:
yamaguch@het.phys.sci.osaka-u.ac.jp}
        
        \\
        \\
        \\
        \llap{$^a$}
        Department of Physics, Osaka University, Toyonaka, Japan
        \\
        \llap{$^b$}
        NIC, DESY Zeuthen, Platanenallee 6, 15738 Zeuthen, Germany
}

\abstract{Atiyah-Singer index theorem on a lattice without boundary is well understood owing to the seminal
work by Hasenfratz {\it et al}.
But its extension to the system with boundary (the so-called Atiyah-Patodi-Singer index theorem),
which plays a crucial role in T-anomaly cancellation between bulk- and edge-modes in
3+1 dimensional topological matters, is known only in the continuum theory and no lattice realization has been made so far.
In this work, we try to non-perturbatively define an alternative index from the lattice domain-wall fermion in 3+1 dimensions. We will show that this new index in the continuum limit, converges
to the Atiyah-Patodi-Singer index defined on a manifold with boundary, which coincides with the
surface of the domain-wall.}

\FullConference{37th International Symposium on Lattice Field Theory - Lattice2019\\
		16-22 June 2019\\
		Wuhan, China}

\begin{document}

\section{Introduction}
The Atiyah-Patodi-Singer(APS) index theorem \cite{Atiyah:1975jf} is an extesion of the Atiyah-Singer(AS) index theorem \cite{Atiyah:1968mp} to a manifold with boundary.
Let us consider a four-dimensional closed Euclidean manifold $X$ with a three-dimensional boundary $Y$.
We assume that $X$ is extending in the region $x_4 > 0$ and the boundary is localized at $x_4 = 0$.
Then the index is given by
\begin{eqnarray}\label{APS index theorem original}
\mathrm{Ind} (D) = \frac{1}{32 \pi^2} \int _{x_4 >0} d^4x \epsilon^{\mu \nu \rho \sigma} F_{\mu \nu} F_{\rho \sigma} - \frac{\eta \left ( i D^{3\mathrm{D}} \right )}{2},
\end{eqnarray}
where $i D^{3 \mathrm{D}}$ is the three-dimensional Dirac operator on $Y$, and $\eta(H)$ is the so-called $\eta$-invariant which is defined by a regularized summation of signs of eigenvalues of $H$:
\begin{eqnarray}
\eta(H) \equiv \sum \mathrm{sgn}(\lambda) = \mathrm{Tr} H/\sqrt{H^2},
\end{eqnarray}
where $\lambda$ is eigenvalues of $H$.
The first term of Eq. (\ref{APS index theorem original}) is equivalent to an integral of the instanton number density, which is not an integer generally on a manifold with boundary.
The second term of (\ref{APS index theorem original}) contains the Chern-Simons term, 
which is not an integer, either.
The APS index theorem claims that the sum of two terms is always an integer.

Recently the APS index theorem is used to the study of condensed matter physics. 
It is related to the physics of $3+1$-dimensional topological insulator with $2+1$-dimensional edge, where the time-reversal(T) symmetry is protected.
The partition function of the edge is proportional to exponential of the $\eta$-invariant of the three-dimensional Dirac operator on the edge:
\begin{eqnarray}
Z_\mathrm{edge} = \mathrm{Det} \left ( i D^{3 \mathrm{D}} \right ) \propto \exp \left [- i \pi \eta\left ( i D^{3 \mathrm{D}} \right ) /2 \right ].
\end{eqnarray}
Similarly the partition function of the bulk is
\begin{eqnarray}
Z_\mathrm{bulk} \propto \exp \left [ i \pi \frac{1}{32\pi^2} \int _{x_4 >0} d^4x \epsilon^{\mu \nu \rho \sigma} F_{\mu \nu } F_{\rho \sigma} \right ].
\end{eqnarray}
As we mentioned, each of these two factors is complex in general, which means that these partition functions break the T-symmetry.
But the total partition function
\begin{eqnarray}
Z_\mathrm{total} \propto \exp \left [ i \pi \frac{1}{32\pi^2} \int _{x_4 >0} d^4x \epsilon^{\mu \nu \rho \sigma} F_{\mu \nu } F_{\rho \sigma} - i \pi \eta\left ( i D^{3 \mathrm{D}} \right ) /2  \right ] = \exp \left [ i \pi \mathrm{Ind} (D) \right ]
\end{eqnarray}
is real, therefore time-reversal symmetry is protected.
Namely, the APS index theorem describes the bulk-edge correspondence in the symmetry protected topological insulator by T-anomaly cancellation between bulk and edge \cite{Witten:2015aba}.

However, the original set-up by APS is quite different from topological insulators.
APS considered a Dirac operator for massless fermions with a non-local boundary condition(APS boundary condition), under which the edge-localized modes are not allowed to exist.
On the other hand, the electron in a topological insulator is massive in the bulk, and the edge-localized modes appear.
It is, therefore, a mathematical puzzle of why the APS index is related to the massive fermion systems.

To fill the gap, three of the authors proposed a new formulation of the APS index theorem \cite{Fukaya:2017tsq} using domain-wall fermion \cite{Kaplan:1992bt}.
Since there exist massive fermions in the bulk and massless edge localized modes on the kink, the domain-wall fermion shares similar properties with the topological insulator.
They showed that in continuum theory the $\eta$-invariant of the domain-wall Dirac operator $H_\mathrm{DW}^c = \gamma_5 (D - M\mathrm{sgn} (x_4) )$ with appropriate regularization(Pauli-Villars(PV) regularization) coincides with the APS formula:
\begin{eqnarray}\label{AS index with eta-inv. continuum theory}
-\frac{1}{2} \eta\left ( H_\mathrm{DW} ^c \right )^{\mathrm{PV} reg.} = \frac{1}{32 \pi^2} \int _{x_4 >0} d^4x \epsilon^{\mu \nu \rho \sigma} F_{\mu \nu} F_{\rho \sigma} - \frac{\eta \left ( i D^{3\mathrm{D}} \right )}{2}.
\end{eqnarray}
This new formulation of the APS index does not require any non-local boundary conditions.
Recently, its mathematical justification was given by \cite{Fukaya:2019qlf}.

In \cite{Fukaya:2017tsq}, they also showed that the AS index is given by a massive Dirac operator:
\begin{eqnarray}
\mathrm{Ind}_\mathrm{AS} (D) = - \frac{1}{2} \eta \left (\gamma_5 (D-M ) \right ) ^{\mathrm{PV}reg.},
\end{eqnarray}
where the Pauli-Villars mass has an opposite sign to $M$.
This fact that the index theorems can be reformulated by massive Dirac operators is quite suggestive since in the reformulation the chiral symmetry is not very important.

Keeping the unimportance of the chiral symmetry in mind, let us revisit the lattice formulation of the Atiyah-Singer index theorem established by Hasenfratz et al. \cite{Hasenfratz:1998ri}.
Using the Dirac operator which satisfies Ginsparg-Wilson(GW) relation $\gamma_5 D + D \gamma_5 = a D \gamma_5 D$ \cite{Ginsparg:1981bj}, the AS index can be given by
\begin{eqnarray}\label{AS index lattice}
\mathrm{Ind}_\mathrm{AS} (D) = \mathrm{Tr} \gamma_5 \left ( 1 - a D /2 \right ),
\end{eqnarray}
where $a$ is lattice spacing.
As an example, let us take the overlap Dirac operator \cite{Neuberger:1997fp}
\begin{eqnarray}\label{overlap Dirac operator}
a D_\mathrm{ov} =   1 + \gamma_5 H_\mathrm{W} / \sqrt{H_\mathrm{W} ^2}, \ \ \ H_\mathrm{W} = \gamma_5 \left ( D_\mathrm{W} - M \right ), \ \ \ M = 1/a,
\end{eqnarray}
where $D_\mathrm{W}$ is the Wilson-Dirac operator.
The convergence of Eq. (\ref{AS index lattice}) with the overlap Dirac operator to the AS index in the continuum limit was confirmed by \cite{Kikukawa:1998pd, Luscher:1998kn, Fujikawa:1998if, Suzuki:1998yz, Adams:1998eg}.

Simply substituting Eq. (\ref{overlap Dirac operator}) into Eq. (\ref{AS index lattice}), we can easily show that the index is equivalent to the $\eta$-invariant of massive Wilson-Dirac operator,
\begin{eqnarray}
\mathrm{Ind}_\mathrm{AS} (D) = - \frac{1}{2} \eta(H_\mathrm{W}) = -\frac{1}{2} \mathrm{Tr}H_\mathrm{W}/\sqrt{H_\mathrm{W}^2},
\end{eqnarray} 
which is a naive lattice discretization of Eq. (\ref{AS index with eta-inv. continuum theory}) by the Wilson-Dirac operator.

The original AS and APS indices require exact chiral symmetry to define the chiral zero modes.
The overlap fermion meets this requirement by the GW relation.
But for our new formulation with the $\eta$-invariant of massive Dirac operator, the chiral symmetry is less important.
The fact that the index of the overlap Dirac operator is the same as the $\eta$-invariant of the massive Wilson-Dirac operator strongly supports a hypothesis that the naive discretization of the $\eta$-invariant of the domain-wall Dirac operator agrees with the APS index in the continuum limit.
As shown in \cite{Fukaya:2019myi}, the original APS boundary condition is difficult to realize on a lattice with the overlap Dirac operator.
Also, any boundary condition would break the GW relation, which makes it impossible to define the APS index on a lattice by the chiral zero modes.

The $\eta$-invariant of the massive Dirac operator, gives a unified view of the index theorems.
In the continuum theory, the APS index theorem is given by just adding a kink structure to the mass in the AS formula.
For the lattice version of the AS index, we only need the Wilson-Dirac operator.
The application to the APS index is therefore straightforward.
Note that $H_\mathrm{DW}$ is a four-dimensional hermitian operator, $\eta( H_\mathrm{DW} ) /2$ is always an integer by definition.
And its variation is always zero under a condition that eigenvalues of $H_\mathrm{DW}$ do not cross zero.

In the following, we will see that the above observation is correct, showing
\begin{eqnarray}
-\frac{1}{2} \eta(H_\mathrm{DW}) = \frac{1}{32 \pi^2} \int _{x_4 >0} d^4x \epsilon^{\mu \nu \rho \sigma} F_{\mu \nu} F_{\rho \sigma} - \frac{\eta \left ( i D^{3\mathrm{D}} \right )}{2},
\end{eqnarray}
in the continuum limit.

\section{Lattice set-up}
We consider the Wilson-Dirac operator with a mass term having a kink structure:
\begin{eqnarray}
  H_{DW} = \gamma_5 \left[D_W-M_1\epsilon\left(x_4+\frac{a}{2}\right)+M_2\right],
\end{eqnarray}
where $\epsilon(x) = \mathrm{sgn}(x)$.
The domain-wall is located at $x_4 = a/2$.
Since the index is defined on a compact manifold, we should consider in a compact space but here we proceed as if we were on an infinite lattice to make the presentation simpler. 
See \cite{Fukaya:2019myi} for more precise treatment.

The key of this work is to find a good complete set to evaluate the $\eta$-invariant as
\begin{eqnarray}\label{evaluation of the eta-inv.}
- \frac{1}{2} \eta ( H_\mathrm{DW} ) = - \frac{1}{2} \mathrm{Tr} \left [H_\mathrm{DW}\sqrt{H_\mathrm{DW}^2}\right ]= \sum_{x, n} \Phi_n ^\dagger (x) \left [H_\mathrm{DW}\sqrt{H_\mathrm{DW}^2}\right ] \Phi_n (x).
\end{eqnarray}
We cannot use simple plane waves due to the loss of translational symmetry in the $x_4$-direction.
For the three directions, which does not have domain-walls, the plane wave set $\psi_p ^{3 \mathrm{D}} = e^{i \bm{p} \cdot \bm{x}} / ( 2\pi)^{3/2}$ is still useful.
We denote the momentum by $\bm{p} = (p_1,p_2,p_3)$ and we assign  two-spinor components to this wave functions.

We consider a complete set given by a direct product $\psi_p^{3\mathrm{D}} \otimes \phi(x_4)$ which is the eigenfunctions of the squared free domain-wall Dirac operator.
Denoting $s_i = \sin (p_i a)$ and $c_i = \cos(p_i a)$, the squared free  domain-wall Dirac operator is expressed by
\begin{eqnarray}
  a^2 (H_{DW}^{\rm 0})^2 &=& s_i^2+ \theta(x_4+a/2)\{M_+^2-(1+M_+)(a^2 \nabla^*_4\nabla_4)\},\nonumber\\
  && + \theta(-x_4-a/2)\{M_-^2-(1+M_-)(a^2 \nabla^*_4\nabla_4)\},\nonumber\\&&
  + 2M_1a (P_+ \delta_{x_4, -a}S_4^+ - P_-\delta_{x_4, 0}S_4^- ),\\
  M_\pm &=& \sum_{i=1,2,3} (1-c_i) \mp M_1a +M_2a,
\end{eqnarray}
where $\theta(x) = \left ( \epsilon(x) +1 \right ) /2$ is the step function, $\nabla_\mu$ and $\nabla_\mu ^*$ denote the forward and backward difference operators respectively, $P_\pm = ( 1+ \gamma_4 )/ 2$ and $S_\mu ^\pm$ is a shift operator operating as $S_\mu ^\pm f(x) = f(x \pm \hat{\mu} a)$.
We have three types of the eigenfunctions in the $x_4$-direction of $a^2 (H_{DW}^{\rm 0})^2$: (1) edge-localized modes at $x_4 =0$, (2) plane wave modes in the region $x_4 \geq 0$, and (3) plane wave modes at any $x_4$.

To simplify the computation, we take the Wilson parameter unity, and take the limit where $M_1 + M_2 \rightarrow \infty$ while the difference is fixed to a finite value $M_1 - M_2 = M > 0$.
After taking this limit, the system is equivalent to the Shamir-type domain-wall fermion \cite{Shamir:1993zy,Furman:1994ky}.
After taking this limit three types of eigenfunctions are reduced to the type (1) and (2) only which still make a complete set.
More explicitly, we have
\begin{eqnarray}
\phi_-^{\rm edge}(x_4) &=& \sqrt{-M_+(2+M_+)/a}e^{-K x_4},\\
\phi_+^\omega(x_4) &=& \frac{1}{\sqrt{2\pi}}(e^{i\omega (x_4+a)}-e^{-i\omega (x_4+a)}),\\
\phi_-^\omega(x_4) &=& \frac{1}{\sqrt{2\pi}}(C_\omega e^{i\omega x_4}-C_\omega^* e^{-i\omega x_4}),
\end{eqnarray}
in the region $x_4\geq 0$, where the subscript $\pm$ denotes the eigenvalue of $\gamma_4=\pm 1$, 
\begin{eqnarray}
  K = - \ln (1+M_+)/a,\hspace{0.5cm}
  C_\omega &=& - \frac{(1+M_+)e^{i\omega a}-1}{|(1+M_+)e^{i\omega a}-1|}.
\end{eqnarray}
Due to the normalizability of the edge-localized modes, we can get the fermion mass condition as $| 1 + M_+ | < 0$ which is equivalent to $0 < Ma < 2$ in the continuum limit.
This condition eliminates the contribution from the doubler modes.

\section{The evaluation of the $\eta$-invariant}
Using the complete set which is derived in the previous section, we can completely decompose the $\eta$-invariant into bulk contribution and edge contribution.
\begin{eqnarray}
- \frac{1}{2} \eta(H_\mathrm{DW} ) = -\frac{1}{2}{\mathrm{Tr}}_\mathrm{bulk} \left [H_{DW}/\sqrt{H_{DW}^2} \right ] -\frac{1}{2}{\mathrm{Tr}}_\mathrm{edge} \left [ H_{DW}\sqrt{H_{DW}^2} \right ].
\end{eqnarray}
\subsection{The evaluation of the bulk contribution}
For the bulk contribution, we consider the density of the $\eta$-invariant:
\begin{eqnarray}
  -\frac{1}{2}{\rm tr} \left [ H_{DW}/\sqrt{H_{DW}^2}\right ](x)^{\rm bulk} 
 &=&
 -\left.
 \frac{1}{2}
  \sum_{g=\pm} \int_0^{\pi/a} d\omega \int_{-\pi/a}^{\pi/a}d^3p
  \right\{\nonumber\\&&\left.
  \hspace{-2.5cm} [\psi^{3 \mathrm{D}}_{p}(\vec{x})\otimes \phi^\omega_g(x_4)]^\dagger
  {\rm tr}\left[P_g \left ( H_{DW}/\sqrt{H_{DW}^2}\right ) P_g\right][\psi^{\rm 3D}_{p}(\vec{x})\otimes\phi^\omega_g(x_4)]\right\}. 
\end{eqnarray}
Here the trace is taken over color and spinor indices only.
We decompose the squared domain-wall Dirac operator into the free part $(H_{DW}^{\rm 0})^2$ and the other part $\Delta H_{DW}^2$ which has the gauge field dependence:
\begin{eqnarray}
  H_{DW}^2 &=& (H_{DW}^{\rm 0})^2 + \Delta H_{DW}^2,\\
  \Delta H_{DW}^2 &=& -\frac{1}{4}\sum_{\mu,\nu}[\gamma^\mu,\gamma^\nu][\tilde{D}_\mu, \tilde{D}_\nu] - \gamma^\mu[\tilde{D}_\mu,\tilde{R}]+ \cdots,  
\end{eqnarray}
where $\cdots$ are the terms having no $\gamma$ structures, and
\begin{eqnarray}
  \tilde{D}_\mu &=& \frac{1}{2a}\left[e^{ip_\mu a}(U_\mu(x) S^+_\mu-1) -e^{-ip_\mu a}(S^-_\mu U_\mu(x)^\dagger -1)\right],
  \\
  \tilde{R} &=& -\frac{1}{2a}\sum_\mu \left[e^{ip_\mu a}(U_\mu(x) S^+_\mu-1)\right. 
    \left.
    + e^{-ip_\mu a}(S^-_\mu U_\mu(x)^\dagger -1)\right],
\end{eqnarray}
assuming them to operate on $[\psi^{3\mathrm{D}}_{p}(\vec{x})\otimes\phi^{ \mathrm{edge}}_-(x_4)]$ and denoting $U_\mu(x)$ as the link variables.

Expanding $1/ \sqrt{H_\mathrm{DW}^2}$ in $\Delta H_\mathrm{DW}^2$ which is higher-order term in $a$, we can show that many terms vanish due to the spinor structure.
The only surviving term is
\begin{eqnarray}
-\frac{1}{2}{\rm tr} \left [ H_{DW}/\sqrt{H_{DW}^2}(x)^{\rm bulk}  \right ] = \left ( I(Ma) + I^\mathrm{DW}(Ma, x_4) \right ) \frac{1}{32\pi^2} \epsilon^{\mu \nu \rho \sigma} \mathrm{tr} F_{\mu \nu} F_{\rho \sigma} 
\end{eqnarray}
upto $\mathcal{O}(a)$ corrections.
The first term is 
\begin{eqnarray}
I(Ma) = \frac{3a^4}{8\pi^2} \int^{\pi/a} _{-\pi/a} d^3 p d \omega \prod_\mu c_\mu \frac{- M_+' + \sum_\nu s_\nu^2 / c_\nu}{\left ( s_\mu^2 + \left [ \sum_\mu (1 - c_\mu ) - Ma\right ]^2 \right ) ^{5/2}},
\end{eqnarray}
which was already evaluated in \cite{Suzuki:1998yz}, and $I(Ma) = 1$ in the continuum limit.
We can also show that the second term is suppressed by a factor of $1/M$.
Since $I(Ma) + I^\mathrm{DW}(Ma, x_4) = 1 + \mathcal{O}(1/M)$, we obtain the standard curvature term from the bulk contribution.

\subsection{The evaluation of the edge contribution}
For the edge contribution, first we consider the domain-wall Dirac operator in $U_4 = 1$ gauge:
\begin{eqnarray}
 a H_\mathrm{DW} = \gamma_5 \left [ -a P_- \nabla_4 + a P_+ \nabla_4^* + a \gamma_i D_i(x_4) + M_+(x_4) \right ].
\end{eqnarray}
Note that $D_i (x_4) $ and $M_+(x_4)$ have $x_4$ dependence through the link variables.
We assume that $x_4$ dependence of link variables is mild.
Then we evaluate the edge contribution in adiabatic approximation ($||U^\dagger \partial_{x_4} U||/ M \ll 1$).

At the leading order of the approximation, eigenfunction of $H_\mathrm{DW}$ is written in $\phi(x) = \phi_{\lambda(0)} ^{3\mathrm{D}} ( \bm{x}) \otimes \phi^\mathrm{edge}_- (x_4)$, 
where $\phi_{\lambda(0)} ^{3\mathrm{D}} ( \bm{x})$ is an eigenfunction of $i \sigma_i D_i (x_4 = 0)$ with the eigenvalue $\lambda(0)$.
$\phi^\mathrm{edge}_- (x_4)$ satisfies $-a P_- \nabla_4 \phi^\mathrm{edge}_- (x_4) = -M_+(0) \phi^\mathrm{edge}_- (x_4)$, 
then the operation of $H_\mathrm{DW}$ to $\phi^\mathrm{edge}_- (x_4)$ is
\begin{eqnarray}
 a H_\mathrm{DW} \phi^\mathrm{edge}_- (x_4) = a  
 \left(
  \begin{array}{cc}
    0 & \\
    & i \sigma_i D_i (0)
  \end{array}
  \right)  \phi^\mathrm{edge}_- (x_4).
\end{eqnarray}
Similarly, we can evaluate the higher order of the adiabatic approximation, and we can conclude that it is suppressed by $1/M
$.

Therefore the edge contribution becomes
\begin{eqnarray}
-\frac{1}{2}{\mathrm{Tr}}_\mathrm{edge} \left [ H_{DW}/\sqrt{H_{DW}^2}\right ] = - \sum_{\lambda(0)} \frac{\mathrm{sgn} \lambda(0)}{2} = - \frac{1}{2} \eta\left ( i \sigma_i D_i\right ) |_{x_4 =0},
\end{eqnarray}
upto $\mathcal{O}(1/M)$ corrections.

\section{Summary}
In this work, we have formulated the Atiyah-Pstodi-Singer index theorem on a lattice.
We have shown that the eta invariant of massive Dirac operator gives a unified view of both the Atiyah-Singer and the APS index theorems in the continuum and lattice theory.
To compute the $\eta$-invariant of the domain-wall Dirac operator, we have obtained a good complete set which consists of bulk plane wave modes and edge-localized modes, in the Shamir-type limit.
We have computed the contribution from bulk and edge separately, 
 then we have confirmed that the $\eta$-invariant of the Wilson-Dirac operator with domain-wall mass converges to the APS index in the continuum limit.

Acknoledgemants: We thank H. Suzuki for his instruction on the computation of $I(M)$.
This work was supported in part by JSPS KAKENHI Grant Number JP15K05054, JP18H01216, JP18H04484, JP18J11457, JP18K03620,
and JP19J20559.
The authors thank the Yukawa Institute for Theoretical Physics at Kyoto University. Discussions during the YITP workshop YITP-T-19-01 on ``Frontiers in Lattice QCD and related topics" were useful to complete this work. T.O. would also like to thank YITP for their kind hospitality during his stay.

\end{document}